\documentclass[aip,jmp,preprint]{revtex4-1}

\begin{document}

\title{Tetrads in Yang-Mills geometrodynamics} %Title of paper

\author{Alcides Garat}
%\thanks{}
\affiliation{1. Instituto de F\'{\i}sica, Facultad de Ciencias, Igu\'a 4225, esq. Mataojo, Montevideo, Uruguay.}

\date{February 10th, 2006}

\begin{abstract}
A new set of tetrads is introduced within the framework of $SU(2) \times U(1)$ Yang-Mills field theories in four dimensional Lorentz curved spacetimes. Each one of these tetrads diagonalizes separately and explicitly each term of the Yang-Mills stress-energy tensor. Therefore, three pairs of planes also known as blades, can be defined, and make up the underlying geometrical structure, at each point. These tetrad vectors are gauge dependent on one hand, and also in their definition, there is an additional inherent freedom in the choice of two vector fields. In order to get rid of the gauge dependence, another set of tetrads is defined, such that the only choice we have to make is for the two vector fields. A particular choice is made for these two vector fields such that they are gauge dependent, but the transformation properties of these tetrads are analogous to those already known for curved spacetimes where only electromagnetic fields are present. This analogy allows to establish group isomorphisms between the local gauge group $SU(2)$, and the tensor product of the groups of local Lorentz tetrad transformations, either on blade one or blade two. These theorems show explicitly that the local internal groups of transformations are isomorphic to local spacetime groups of transformations. As an example of application of these new tetrads, we exhibit three new gauge invariant objects, and using these objects we show how to diagonalize the Yang-Mills stress-energy tensor in a gauge invariant way.
\end{abstract}


\maketitle %\maketitle must follow title, authors, abstract and \pacs

\section{Introduction}
\label{introduction}
Electromagnetic fields can be used to introduce at each point in a four dimensional Lorentz spacetime
a local structure where a pair of blades can be defined through the use of a special tetrad \cite{A}$^{,}$\cite{MW}$^{,}$\cite{SCH}$^{,}$\cite{MTW}. Schouten defined what he called, a two-bladed structure in a spacetime \cite{SCH}. These blades are planes, defined by the tetrad vectors. The timelike and one spacelike vectors define what we called in \cite{A}, blade one, and the other two spacelike vectors, define blade two. In turn, this tetrad is built out of the extremal field, defined through a duality rotation of the local electromagnetic field. At every point in spacetime there is a duality rotation by an angle $-\alpha$ that transforms a non-null electromagnetic field into an extremal field,

\begin{equation}
\xi_{\mu\nu} = e^{-\ast \alpha} f_{\mu\nu}\ .\label{dref}
\end{equation}

Extremal fields are essentially electric fields and they satisfy,

\begin{equation}
\xi_{\mu\nu} \ast \xi^{\mu\nu}= 0\ .
\end{equation}

In \cite{A} it was proved that the local electromagnetic gauge group is isomorphic to the local group of Lorentz transformations of the tetrad vectors on blades one and two. The relation between the electromagnetic gauge transformations and the local Lorentz tetrad transformations on both blades was straightforward. The simplification in the expression of all relevant fields and equations was maximum. It was natural then, to ask if similar structures could be built for non-Abelian fields \cite{YM}$^{,}$\cite{RU}$^{,}$\cite{DT}. The Abelian nature of the electromagnetic field results in the existence of just one extremal field and complexion scalar, the non-Abelian nature of a $SU(2)$ field should provide with more geometrical structure and therefore more than one extremal field and complexion. In fact, three extremal fields and three complexions can be defined out of the three $SU(2)$ fields, one per internal index. This is because it is possible to diagonalize each one of the three terms in the Yang-Mills stress-energy tensor separately, and explicitly. It is as if the three $SU(2)$ field components remain algebraically ``decoupled''; but at a price. The tetrad vectors are on one hand $SU(2)$ gauge dependent, and on the other hand, in their definition, there is an inherent freedom in the choice of two vector fields $X^{\mu}$, $Y^{\mu}$. If we transform these two vectors as $X^{\mu} \rightarrow X^{\mu} + \Lambda^{\mu}$, with $\Lambda$ a scalar function, and, $Y^{\mu} \rightarrow Y^{\mu} + \ast \Lambda^{\mu}$, with $\ast \Lambda$ another scalar function, then the tetrad vectors transform in an analogous fashion to the electromagnetic Abelian case \cite{A}, under $U(1)$ transformations. $\Lambda^{\mu}$ is simplifying notation for $\Lambda^{,\mu}$. The problem dwells in the $SU(2)$ local gauge transformations, and the $SU(2)$ local gauge dependence of the ``decoupled'' tetrads. We would like to find a tetrad such that the transformation properties of the tetrad vectors are analogous to the Abelian case, but under $SU(2)$ transformations, that is, the two vectors that define blade one, remain on blade one after the $SU(2)$ transformation, and the two that define blade two, remain on blade two after the $SU(2)$ transformation.
The question presents itself on the reason for asking such a transformation property to be fulfilled by the tetrad vectors. The answer is that the metric tensor once is built out of the tetrad vectors, must be manifestly invariant under $SU(2)$ local gauge transformations. The requirement that for each tetrad, and under local $SU(2)$ gauge transformations, the normalized vectors that define blades one and two should remain on their respective blades, ensures explicitly the invariance of the metric tensor under local $SU(2)$ gauge transformations. In order to find this new tetrad with the required $SU(2)$ gauge transformation properties we proceed to build a new kind of extremal field. In addition, we use a new duality rotation that involves a new complexion, which is in turn invariant under local $SU(2)$ gauge transformations. Once we have this new extremal field, building the new tetrad with the acceptable $SU(2)$ gauge transformation properties is automatic. We are going to name LB1 the group of Lorentz tetrad transformations on blade one. Analogously we name the group of rotations on blade two, LB2. Following the ideas provided in \cite{A} for the Abelian case, as a general guide, it is found that a $SU(2)$ local gauge transformation generates the composition of two transformations. A tetrad transformation, generated by a locally inertial Lorentz coordinate transformation, and a local Lorentz LB1 transformation of the tetrad vectors on blade one. Then, following again the steps in \cite{A} it is proved an isomorphism between the $SU(2)$ group of transformations, and the tensor product of three LB1 groups associated to three different sets of tetrads. As one of the transformations is generated by the group of locally inertial Lorentz coordinate transformations, the non-commutativity of the image is assured. A similar result for blade two with tensor product of three LB2 groups. Through these group isomorphisms between local groups of transformations, we analyze the connection between the gauge and geometrical structures. These theorems in addition to the ones in the Abelian case, prove an explicit isomorphic relation between local groups of ``internal'' transformations and local groups of ``spacetime'' transformations \cite{SWNG}$^{,}$\cite{LORNG}$^{,}$\cite{CMNG}$^{,}$\cite{MK}. They also prove isomorphisms between compact groups of local transformations on one hand, and non-compact groups of local transformations plus discrete transformations on the other hand. As applications of these new tetrads we present three new gauge invariant quantities built out of the stress-energy components, and a new gauge invariant method to diagonalize the Yang-Mills stress-energy tensor.

This manuscript is organized as follows. In section \ref{3complexions}, a set of three tetrads is introduced by just studying the diagonalization of each term in the stress-energy tensor. In section \ref{extremalsu2} new tetrads are introduced such that their transformation properties under $SU(2)$ local gauge transformations follow the same geometrical pattern than the Abelian electromagnetic ones. In section \ref{gaugegeometry}, the transformation properties of the tetrads introduced in the previous section are analyzed, as well as the group isomorphisms associated with them. In section \ref{appli} we introduce three new gauge invariant quantities built out of the stress-energy components, and a new gauge invariant method to diagonalize the Yang-Mills stress-energy tensor. Throughout the paper we use the conventions of \cite{A}$^{,}$\cite{MW}. In particular we use a metric with sign conventions -+++, and $f^{k}_{\mu\nu}$ are the geometrized Yang-Mills field components, $f^{k}_{\mu\nu}= (G^{1/2} / c^2) \: F^{k}_{\mu\nu}$.

\section{Complexions for the decoupled tetrad}
\label{3complexions}

The stress-energy tensor for the $SU(2)$ Yang-Mills field can be written as \cite{MC},

\begin{equation}
T_{\mu\nu}= f^{k} _{\mu\lambda}\:\:f_{\:\:\nu}^{k\:\:\lambda}
+ \ast f^{k}_{\mu\lambda}\:\ast f_{\:\:\nu}^{k\:\:\lambda}\ ,\label{TEM}
\end{equation}

where the summation convention on the internal index $k$ is applied, and
$\ast f^{k}_{\mu\nu}=
{1 \over 2}\:\epsilon_{\mu\nu\sigma\tau}\:f^{k\:\sigma\tau}$
is the dual tensor of $f^{k}_{\mu\nu}$. The duality rotation given by equation (59) in\cite{MW}, can be written separately for each internal index $k$ as,

\begin{equation}
f^{(k)}_{\mu\nu} = \cos\alpha_{k} \:\: \hat{\xi}^{(k)}_{\mu\nu}+
\sin\alpha_{k} \:\: \ast\hat{\xi}^{(k)}_{\mu\nu} \ ,\label{drk}
\end{equation}

or,

\begin{equation}
\hat{\xi}^{(k)}_{\mu\nu} = \cos\alpha_{k} \:\: f^{(k)}_{\mu\nu}-
\sin\alpha_{k} \:\: \ast f^{(k)}_{\mu\nu} \ ,\label{idrk}
\end{equation}

where the summation convention for the index $(k)$ between parenthesis is not applied.
We can also follow the same procedure as in \cite{MW} for each internal index $k$ and define the three complexions by imposing,

\begin{eqnarray}
\hat{\xi}^{(k)}_{\mu\nu}\:\ast \hat{\xi}^{(k)\mu\nu} &=& 0\ .\label{cc}
\end{eqnarray}

As a result,

\begin{eqnarray}
\tan(2\alpha_{k}) = - f^{(k)}_{\mu\nu}\:\ast f^{(k)\mu\nu} / f^{(k)}_{\lambda\rho}\:f^{(k)\lambda\rho}\ .\label{compk}
\end{eqnarray}

Then, it is straightforward to express the stress-energy tensor in terms of the extremal field ``decoupled'' internal components,

\begin{equation}
T_{\mu\nu}= \sum_{k=1}^{3}\: \left( \hat{\xi}^{(k)}_{\mu\lambda}\:\:\hat{\xi}_{\:\:\:\:\:\nu}^{(k)\:\:\lambda}
+ \ast \hat{\xi}^{(k)}_{\mu\lambda} \:\ast \hat{\xi}_{\:\:\:\:\:\nu}^{(k)\:\:\lambda} \right) = \sum_{k=1}^{3}\:T^{(k)}_{\mu\nu} \ .\label{TEMDR}
\end{equation}

Following the Abelian ideas we can define as many sets of tetrad vectors at every point in spacetime, as generators has the gauge group,

\begin{eqnarray}
V_{(1)}^{(k)\:\mu} &=& \hat{\xi}^{(k)\:\mu\lambda}\:\hat{\xi}^{(k)}_{\rho\lambda}\:X^{\rho}
\label{V1}\\
V_{(2)}^{(k)\:\mu} &=& \sqrt{-Q^{(k)}/2} \: \hat{\xi}^{(k)\:\mu\lambda} \: X_{\lambda}
\label{V2}\\
V_{(3)}^{(k)\:\mu} &=& \sqrt{-Q^{(k)}/2} \: \ast \hat{\xi}^{(k)\:\mu\lambda} \: Y_{\lambda}
\label{V3}\\
V_{(4)}^{(k)\:\mu} &=& \ast \hat{\xi}^{(k)\:\mu\lambda}\: \ast\hat{\xi}^{(k)}_{\rho\lambda}
\:Y^{\rho}\ ,\label{V4}
\end{eqnarray}

where $Q^{(k)}=\hat{\xi}^{(k)}_{\mu\nu}\:\hat{\xi}^{(k)\mu\nu}$.
$Q^{(k)}$ is assumed not to be zero.
We are free to choose the vector fields $X^{\lambda}$ and $Y^{\lambda}$, as
long as the four vector fields (\ref{V1}-\ref{V4}) are not trivial.
Two identities in the extremal field are going to be used extensively
in this work, in particular, to prove that tetrad (\ref{V1}-\ref{V4})
diagonalizes the stress-energy tensor $k$ component.
Using the general identity for two antisymmetrical fields,

\begin{eqnarray}
A_{\mu\sigma}\:B^{\nu\sigma} -
\ast B_{\mu\sigma}\: \ast A^{\nu\sigma} &=& \frac{1}{2}
\: \delta_{\mu}^{\:\:\:\nu}\ A_{\sigma\tau}\:B^{\sigma\tau}\ ,\label{idenanti}
\end{eqnarray}

the first identity results from equation (\ref{cc}), which is in fact the algebraic equation or condition imposed, in order to define the $\alpha_{k}$ complexion, and equation (\ref{idenanti}),

\begin{eqnarray}
\hat{\xi}^{(k)}_{\sigma\mu}\:\ast \hat{\xi}^{(k)\mu\nu} &=& 0\ .\label{i1}
\end{eqnarray}

Using again (\ref{idenanti}), we can find the second identity for each internal value of $k$,

\begin{eqnarray}
\hat{\xi}^{(k)}_{\mu\sigma}\:\hat{\xi}^{(k)\nu\sigma} -
\ast \hat{\xi}^{(k)}_{\mu\sigma}\: \ast \hat{\xi}^{(k)\nu\sigma} &=& \frac{1}{2}
\: \delta_{\mu}^{\:\:\:\nu}\ \hat{\xi}^{(k)}_{\sigma\tau}\hat{\xi}^{(k)\sigma\tau}\ .\label{i2}
\end{eqnarray}

When we make iterative use of (\ref{i1}) and (\ref{i2}) we find,

\begin{eqnarray}
V_{(1)}^{(k)\sigma}\:T_{\:\:\:\sigma}^{(k)\:\tau} &=& \frac{Q^{(k)}}{2}\:V_{(1)}^{(k)\tau}
\label{EV1}\\
V_{(2)}^{(k)\sigma}\:T_{\:\:\:\sigma}^{(k)\:\tau} &=& \frac{Q^{(k)}}{2}\:V_{(2)}^{(k)\tau}
\label{EV2}\\
V_{(3)}^{(k)\sigma}\:T_{\:\:\:\sigma}^{(k)\:\tau} &=& -\frac{Q^{(k)}}{2}\:V_{(3)}^{(k)\tau}
\label{EV3}\\
V_{(4)}^{(k)\sigma}\:T_{\:\:\:\sigma}^{(k)\:\tau} &=& -\frac{Q^{(k)}}{2}\:V_{(4)}^{(k)\tau}\ .
\label{EV4}
\end{eqnarray}

In \cite{MW} the stress-energy tensor for the Abelian field was diagonalized through the use of a Minkowskian frame in which the equation for this tensor
was given in equations (34) and (38).
In this work, for non-Abelian fields we provide the explicit expression for the
tetrad in which the stress-energy tensor $k$ component is diagonal.
The freedom we have to choose the vector fields $X^{\mu}$ and $Y^{\mu}$,
represents available freedom that we have to choose the tetrad.
If we make use of equations (\ref{i1}) and (\ref{i2}),
it is straightforward to prove that (\ref{V1}-\ref{V4})
is a set of orthogonal vectors. If transformations of the vector field
$X^{\mu}\rightarrow X^{\mu}+\Lambda^{\mu}$, with $\Lambda$ a scalar function, are introduced in an analogous fashion to \cite{A}, regarding these transformations in the general sense explained in \cite{A}, in the section ``general tetrad'', then, we can carry over into each of the ``decoupled'' tetrads, the same conclusions reached in \cite{A} regarding isomorphisms, for instance. Even though these transformations do not work exactly as in the Abelian case, because the ``decoupled'' fields $\hat{\xi}^{(k)\:\mu\nu}$ are not considered to transform, as transformations that affect only the vector field $X^{\mu}$, they behave as in the Abelian case. The problem arises when we perform local $SU(2)$ gauge transformations in which case the ``decoupled'' fields $\xi^{(k)\mu\nu}$ do transform spoiling the invariance of the metric tensor. The ``decoupled'' tetrads are fundamentally providing information about the number of independent pairs of blades we can build at each point, but their transformation properties do not satisfy the requirement for the invariance of the metric tensor under $SU(2)$ local gauge transformations.

\section{Extremal field in $SU(2)$ geometrodynamics}
\label{extremalsu2}

One of our goals is to build a ``non-decoupled'' tetrad such that $SU(2)$ generated tetrad transformations on blade one, leave the tetrad vectors that generate this blade, on blade one, and similarly for $SU(2)$ generated rotations on blade two. This property is fundamental to ensure the invariance of the metric tensor under $SU(2)$ local gauge transformations, and is going to be used when proving the existence of morphisms between the local $SU(2)$ gauge group and the local LB1, LB2 groups. The ``decoupled'' tetrads clearly do not have this property. The ``decoupled'' tetrad vectors might leave blades one and two after a gauge transformation. Let us define then, a ``non-decoupled'' extremal field as,
\begin{equation}
\zeta_{\mu\nu} = \cos\beta \:\: f_{\mu\nu}-
\sin\beta \:\: \ast f_{\mu\nu} \ ,\label{exsu2}
\end{equation}

In order to define the complexion $\beta$, we are going to impose the $SU(2)$ invariant condition,

\begin{eqnarray}
Tr[\zeta_{\mu\nu}\:\ast \zeta^{\mu\nu}]=\zeta^{k}_{\mu\nu}\:\ast \zeta^{k\mu\nu} &=& 0\ ,\label{ccsu2}
\end{eqnarray}

where the summation convention was applied on the internal index $k$. The complexion condition (\ref{ccsu2}) is not an additional condition for the field strength. We are just using a generalized duality transformation, and defining through it this new local scalar complexion $\beta$. After the fields are available from the equations, not before. We simply generalized the definition for the Abelian complexion, found through a duality transformation as well. Then, the local $SU(2)$ invariant complexion $\beta$ turns out to be,

\begin{eqnarray}
\tan(2\beta) = - f^{k}_{\mu\nu}\:\ast f^{k\mu\nu} / f^{p}_{\lambda\rho}\:f^{p\lambda\rho}\ ,\label{compksu2}
\end{eqnarray}

where again the summation convention was applied on both $k$ and $p$.

 Now we would like to consider gauge covariant derivatives. For instance, the gauge covariant derivatives of the three ``non-decoupled'' extremal field internal components,

\begin{eqnarray}
\zeta_{k\mu\nu\mid\rho} = \zeta_{k\mu\nu\, ; \, \rho} + g \: \epsilon_{klp}\: A_{l\rho}\:\zeta_{p\mu\nu}\ .\label{gcd}
\end{eqnarray}

where $\epsilon_{klp}$ is the completely skew-symmetric tensor in three dimensions with $\epsilon_{123} = 1$, and $g$ is the coupling constant. The symbol ``;'' stands for the usual covariant derivative associated with the metric tensor $g_{\mu\nu}$. If we consider for instance the Einstein-Maxwell-Yang-Mills vacuum field equations,

\begin{eqnarray}
R_{\mu\nu} &=& T^{(ym)}_{\mu\nu} + T^{(em)}_{\mu\nu}\label{eyme}\\
f^{\mu\nu}_{\:\:\:\:\:;\nu} &=& 0 \label{EM1}\\
\ast f^{\mu\nu}_{\:\:\:\:\:;\nu} &=& 0 \label{EM2}\\
f^{k\mu\nu}_{\:\:\:\:\:\:\:\:\mid \nu} &=& 0 \label{ymvfe1}\\
\ast f^{k\mu\nu}_{\:\:\:\:\:\:\:\:\mid \nu} &=& 0 \ . \label{ymvfe2}
\end{eqnarray}

The field equations (\ref{EM1}-\ref{EM2}) provide a hint about the existence of two electromagnetic field potentials, as said in the first paper ``Tetrads in geometrodynamics'', not independent from each other, but due to the symmetry of the equations, available for our construction. $A^{\mu}$ and $\ast A^{\mu}$ are the two electromagnetic potentials. $\ast A^{\mu}$ is therefore a name, we are not using the Hodge map at all in this case. These two potentials are not independent from each other, nonetheless they exist and are available for our construction. Similar for the two Non-Abelian  equations (\ref{ymvfe1}-\ref{ymvfe2}). The Non-Abelian potential $A^{k\mu}$ is available for our construction as well.
With all these elements, we can proceed as an example, to define the antisymmetric field,

\begin{eqnarray}
\omega_{\mu\nu} = Z_{\mu\nu\sigma\tau} \: U^{\sigma\tau}\ ,\label{anti1}
\end{eqnarray}

where $Z_{\mu\nu\sigma\tau}$ could be for instance $\:\zeta^{p}_{\:\sigma\tau}\: \zeta^{p}_{\mu\nu}$, $\:\ast \zeta^{p}_{\:\sigma\tau}\: \ast\zeta^{p}_{\mu\nu}$, $\:\zeta^{p}_{\:\sigma\mu}\: \zeta^{p}_{\tau\nu} - \zeta^{p}_{\:\tau\mu}\: \zeta^{p}_{\sigma\nu}$, $\:\ast\zeta^{p}_{\:\sigma\tau}\: \zeta^{p}_{\mu\nu} + \zeta^{p}_{\:\sigma\tau}\: \ast\zeta^{p}_{\mu\nu}$, $\:f^{p}_{\:\sigma\tau}\: f^{p}_{\mu\nu}$, $\:\ast f^{p}_{\:\sigma\tau}\: \ast f^{p}_{\mu\nu}$, etc. It could also be the standard Riemann tensor $R_{\mu\nu\sigma\tau}$, the Weyl tensor $C_{\mu\nu\sigma\tau}$, the tensor $(g_{\mu\sigma}\:g_{\nu\tau} - g_{\mu\tau}\:g_{\nu\sigma})$ or $R_{\mu\nu\rho\lambda}\:R_{\sigma\tau}^{\:\:\:\:\:\:\rho\lambda}$, etc.
In the case of $U^{\sigma\tau}$ we could have chosen for instance, $\zeta^{k\sigma\rho}_{\:\:\:\:\:\:\:\:\mid\rho}\:\ast \zeta^{k\tau\lambda}_{\:\:\:\:\:\:\:\:\mid\lambda}$, or $\left(\zeta^{k\sigma\rho}\:\ast \zeta^{k\tau\lambda}-\ast\zeta^{k\sigma\rho}\:\zeta^{k\tau\lambda}\right)\:T_{\rho\lambda}$, etc. The summation convention on the internal index $k$ as well as $p$ was applied. It is clear that (\ref{anti1}) is invariant under $SU(2)$ local gauge transformations. Expression (\ref{anti1}) is nothing but an explicit example among many. If our choice for an antisymmetric field is (\ref{anti1}), then the duality rotation we perform next, in order to obtain the new extremal field, is the duality rotation that we have available on the $Z_{\mu\nu\sigma\tau}$ tensor,

\begin{eqnarray}
\epsilon_{\mu\nu} = \cos\vartheta \: \omega_{\mu\nu} - \sin\vartheta \:
\ast \omega_{\mu\nu}\ .\label{extremalR}
\end{eqnarray}

As always we choose this complexion $\vartheta$ to be defined by the condition,

\begin{eqnarray}
\epsilon_{\mu\nu}\:\ast \epsilon^{\mu\nu} &=& 0\ ,\label{rc}
\end{eqnarray}

which implies that,

\begin{eqnarray}
\tan(2\vartheta) = - \omega_{\mu\nu}\:\ast \omega^{\mu\nu} / \omega_{\lambda\rho}\:\omega^{\lambda\rho}\ .\label{compr}
\end{eqnarray}

This new kind of local $SU(2)$ gauge invariant extremal tensor $\epsilon_{\mu\nu}$, allows in turn for the construction of the new tetrad,

\begin{eqnarray}
S_{(1)}^{\mu} &=& \epsilon^{\mu\lambda}\:\epsilon_{\rho\lambda}\:X^{\rho}
\label{S1}\\
S_{(2)}^{\mu} &=& \sqrt{-Q_{ym}/2} \: \epsilon^{\mu\lambda} \: X_{\lambda}
\label{S2}\\
S_{(3)}^{\mu} &=& \sqrt{-Q_{ym}/2} \: \ast \epsilon^{\mu\lambda} \: Y_{\lambda}
\label{S3}\\
S_{(4)}^{\mu} &=& \ast \epsilon^{\mu\lambda}\: \ast\epsilon_{\rho\lambda}
\:Y^{\rho}\ ,\label{S4}
\end{eqnarray}

where $Q_{ym} = \epsilon_{\mu\nu}\:\epsilon^{\mu\nu}$. It is straightforward using (\ref{idenanti}), to prove that they are orthogonal. We are going to call for future reference for instance $\epsilon^{\mu\lambda}\:\epsilon_{\rho\lambda}$ the skeleton of the tetrad vector $S_{(1)}^{\mu}$, and $X^{\rho}$ the gauge vector. In the case of $S_{(3)}^{\mu}$, the skeleton will be $\ast \epsilon^{\mu\lambda}$, and $Y_{\lambda}$ will be the gauge vector. It is clear now that skeletons are gauge invariant. This property guarantees that the vectors under local $U(1)$ or $SU(2)$ gauge transformations are not going to leave their original planes or blades, keeping therefore the metric tensor explicitly invariant.

\section{gauge geometry}
\label{gaugegeometry}

The question remains about the choice that we can make for the two gauge vector fields
$X^{\sigma}$ and $Y^{\sigma}$ in (\ref{S1}-\ref{S4}) such that we can reproduce in the $SU(2)$ environment, the tetrad transformation properties of the Abelian environment.

One possible choice could be $X^{\sigma} = Y^{\sigma} = Tr[\Sigma^{\alpha\beta}\:E_{\alpha}^{\:\:\rho}\: E_{\beta}^{\:\:\lambda}\:\ast \xi_{\rho}^{\:\:\sigma}\:\ast \xi_{\lambda\tau}\:A^{\tau}]$. The nature of the object $\Sigma^{\alpha\beta}$ is explained in section \ref{sec:appII}. $E_{\alpha}^{\:\:\rho}$ are tetrad vectors that transform from a locally inertial coordinate system, into a general curvilinear coordinate system. Greek indices $\alpha$, $\beta$, $\delta$, $\epsilon$, $\gamma$, and $\kappa$, are reserved for locally inertial coordinate systems. There is a particular explicit choice that we can make for these tetrads $E_{\alpha}^{\:\:\rho}$. We can choose the tetrad vectors we already know from \cite{A}, for electromagnetic fields in curved space-times. Following the same notation in \cite{A}, we call $E_{o}^{\:\:\rho} = U^{\rho}$, $E_{1}^{\:\:\rho} = V^{\rho}$, $E_{2}^{\:\:\rho} = Z^{\rho}$, $E_{3}^{\:\:\rho} = W^{\rho}$. The electromagnetic extremal tensor $\xi_{\rho\sigma}$, and its dual $\ast \xi_{\rho\sigma}$ are also already known from \cite{A}. That is, we are making use of the already defined tetrads built for space-times where electromagnetic fields are present, in order to allow for the use of the object $\Sigma^{\alpha\beta}$ which is key in our construction. The key lies in the translating quality of this object between $SU(2)$ local gauge transformations and local Lorentz transformations. We would like to consider one more property of these chosen vector fields $X^{\rho}$ and $Y^{\rho}$. The structure $E_{\alpha}^{\:\:[\rho}\:E_{\beta}^{\:\:\lambda]}\:\ast \xi_{\rho\sigma}\:\ast \xi_{\lambda\tau}$ is invariant under $U(1)$ local gauge transformations. Essentially, because of the electromagnetic extremal field property \cite{A}$^{,}$\cite{MW}, $\xi_{\mu\sigma}\:\ast \xi^{\mu\tau} = 0$.

%$E_{\alpha}^{\:\:[\rho}\:E_{\beta}^{\:\:\lambda]}\:\ast \xi_{\rho\sigma}\:\ast \xi_{\lambda\tau}$

Along the lines established in \cite{A} we can study the $SU(2)$ gauge transformation properties of these two vector fields. We observe that under local $SU(2)$ gauge transformations $S$, introduced in section \ref{sec:appI},

\begin{eqnarray}
A_{\mu} \: \rightarrow S^{-1}\:A_{\mu}\:S + {\imath \over g} S^{-1}\:\partial_{\mu}(S) \label{GTA}
\end{eqnarray}

while the gauge vectors transform as,

\begin{eqnarray}
Tr[\Sigma^{\alpha\beta}\:E_{\alpha}^{\:\:\rho}\: E_{\beta}^{\:\:\lambda}\:\ast \xi_{\rho\sigma}\:\ast \xi_{\lambda\tau}\:A^{\tau}] \: \rightarrow Tr[\Sigma^{\alpha\beta}\:E_{\alpha}^{\:\:\rho}\: E_{\beta}^{\:\:\lambda}\:\ast \xi_{\rho\sigma}\:\ast \xi_{\lambda\tau}\:S^{-1}\:A^{\tau}\:S] + \nonumber \\
{\imath \over g}\: Tr[\Sigma^{\alpha\beta}\:E_{\alpha}^{\:\:\rho}\: E_{\beta}^{\:\:\lambda}\:\ast \xi_{\rho\sigma}\:\ast \xi_{\lambda\tau}\:S^{-1}\:\partial^{\tau}\:(S)] \label{XCHOICE}
\end{eqnarray}

The field strength transforms as usual, $f_{\mu\nu} \rightarrow S^{-1}\:f_{\mu\nu}\:S$, and similar for the extremal $\xi_{\mu\nu}$, and their duals. Then, we can follow exactly the same guidelines laid out in \cite{A}, to study the gauge transformations of the tetrad vectors on blades one, and two. We can carry over into the present work, all the analysis done in the gauge geometry section in \cite{A}. It is clear that the vectors $S_{(1)}^{\mu}$ and $S_{(2)}^{\mu}$, by virtue of their own construction, remain on blade one after the (\ref{GTA}) transformation. It is also evident that after the transformation they are orthogonal. These two facts mean that the metric tensor is invariant under the transformations (\ref{GTA}) when the two vectors are normalized. We are assumming for simplicity that $S_{(1)}^{\mu}$ is timelike and $S_{(2)}^{\mu}$ spacelike, both vectors non-trivial. Let us study then, the transformation of vectors (\ref{S1}-\ref{S2}) under the transformations (\ref{GTA}). We can write,

\begin{eqnarray}
\tilde{S}_{(1)}^{\mu} = \epsilon^{\mu\nu}\:\epsilon^{\sigma}_{\:\:\nu}\: Tr[\Sigma^{\alpha\beta}\:E_{\alpha}^{\:\:\rho}\: E_{\beta}^{\:\:\lambda}\:\ast \xi_{\rho\sigma}\:\ast \xi_{\lambda\tau}\:S^{-1}\:A^{\tau}\:S] + \nonumber \\
{\imath \over g} \:\epsilon^{\mu\nu}\:\epsilon^{\sigma}_{\:\:\nu}\: Tr[\Sigma^{\alpha\beta}\:E_{\alpha}^{\:\:\rho}\: E_{\beta}^{\:\:\lambda}\:\ast \xi_{\rho\sigma}\:\ast \xi_{\lambda\tau}\:S^{-1}\:\partial^{\tau}\:(S)]
\label{S1T}
\end{eqnarray}
\begin{eqnarray}
\tilde{S}_{(2)}^{\mu} = \sqrt{-Q_{ym}/2} \:\left( \epsilon^{\mu\sigma} \: Tr[\Sigma^{\alpha\beta}\:E_{\alpha}^{\:\:\rho}\: E_{\beta}^{\:\:\lambda}\:\ast \xi_{\rho\sigma}\:\ast \xi_{\lambda\tau}\:S^{-1}\:A^{\tau}\:S] \right. + \nonumber \\ \left.
{\imath \over g}\: \epsilon^{\mu\sigma} \: Tr[\Sigma^{\alpha\beta}\:E_{\alpha}^{\:\:\rho}\: E_{\beta}^{\:\:\lambda}\:\ast \xi_{\rho\sigma}\:\ast \xi_{\lambda\tau}\:S^{-1}\:\partial^{\tau}\:(S)] \right) \ .
\label{S2T}
\end{eqnarray}

It is possible to rewrite equations (\ref{S1T}-\ref{S2T}) as,

\begin{eqnarray}
\tilde{S}_{(1)}^{\mu} = \epsilon^{\mu\nu}\:\epsilon^{\sigma}_{\:\:\nu}\: Tr[S\:\Sigma^{\alpha\beta}\:S^{-1}\:E_{\alpha}^{\:\:\rho}\: E_{\beta}^{\:\:\lambda}\:\ast \xi_{\rho\sigma}\:\ast \xi_{\lambda\tau}\:A^{\tau}] + \nonumber \\ {\imath \over g} \:\epsilon^{\mu\nu}\:\epsilon^{\sigma}_{\:\:\nu}\: Tr[S\:\Sigma^{\alpha\beta}\:S^{-1}\:E_{\alpha}^{\:\:\rho}\: E_{\beta}^{\:\:\lambda}\:\ast \xi_{\rho\sigma}\:\ast \xi_{\lambda\tau}\:\partial^{\tau}\:(S)\:S^{-1}]
\label{S1R}
\end{eqnarray}
\begin{eqnarray}
\tilde{S}_{(2)}^{\mu} = \sqrt{-Q_{ym}/2} \: \left( \epsilon^{\mu\sigma} \: Tr[S\:\Sigma^{\alpha\beta}\:S^{-1}\:E_{\alpha}^{\:\:\rho}\: E_{\beta}^{\:\:\lambda}\:\ast \xi_{\rho\sigma}\:\ast \xi_{\lambda\tau}\:A^{\tau}] \right. + \nonumber \\ \left.
{\imath \over g}\: \epsilon^{\mu\sigma} \: Tr[S\:\Sigma^{\alpha\beta}\:S^{-1}\:E_{\alpha}^{\:\:\rho}\: E_{\beta}^{\:\:\lambda}\:\ast \xi_{\rho\sigma}\:\ast \xi_{\lambda\tau}\:\partial^{\tau}\:(S)\:S^{-1}] \right) \ .
\label{S2R}
\end{eqnarray}

For the sake of simplicity we are using the notation, $\Lambda^{(-1)\,\alpha}_{\:\:\:\:\:\:\:\:\:\:\:\:\delta} = \tilde{\Lambda}^{\alpha}_{\:\:\:\delta}$, and no confusion should arise with the transformed vectors $\tilde{S}_{(1)}^{\mu}$, $\:\tilde{S}_{(2)}^{\mu}$, for instance. Now, we can make use of the local transformation properties of the objects $\Sigma^{\alpha\beta}$, see section \ref{sec:appII}, and write,

\begin{eqnarray}
\tilde{S}_{(1)}^{\mu} = \epsilon^{\mu\nu}\:\epsilon^{\sigma}_{\:\:\nu}\: Tr[\tilde{\Lambda}^{\alpha}_{\:\:\:\delta}\:\tilde{\Lambda}^{\beta}_{\:\:\:\gamma}\:\Sigma^{\delta\gamma}\:E_{\alpha}^{\:\:\rho}\: E_{\beta}^{\:\:\lambda}\:\ast \xi_{\rho\sigma}\:\ast \xi_{\lambda\tau}\:A^{\tau}] + \nonumber \\ {\imath \over g} \:\epsilon^{\mu\nu}\:\epsilon^{\sigma}_{\:\:\nu}\: Tr[\tilde{\Lambda}^{\alpha}_{\:\:\:\delta}\:\tilde{\Lambda}^{\beta}_{\:\:\:\gamma}\:\Sigma^{\delta\gamma}\:E_{\alpha}^{\:\:\rho}\: E_{\beta}^{\:\:\lambda}\:\ast \xi_{\rho\sigma}\:\ast \xi_{\lambda\tau}\:\partial^{\tau}\:(S)\:S^{-1}]
\label{S1L}
\end{eqnarray}
\begin{eqnarray}
\tilde{S}_{(2)}^{\mu} = \sqrt{-Q_{ym}/2} \:\left( \epsilon^{\mu\sigma} \: Tr[\tilde{\Lambda}^{\alpha}_{\:\:\:\delta}\:\tilde{\Lambda}^{\beta}_{\:\:\:\gamma}\:\Sigma^{\delta\gamma}\:E_{\alpha}^{\:\:\rho}\: E_{\beta}^{\:\:\lambda}\:\ast \xi_{\rho\sigma}\:\ast \xi_{\lambda\tau}\:A^{\tau}] \right. + \nonumber \\ \left. {\imath \over g}\:\epsilon^{\mu\sigma} \: Tr[\tilde{\Lambda}^{\alpha}_{\:\:\:\delta}\:\tilde{\Lambda}^{\beta}_{\:\:\:\gamma}\:\Sigma^{\delta\gamma}\:E_{\alpha}^{\:\:\rho}\: E_{\beta}^{\:\:\lambda}\:\ast \xi_{\rho\sigma}\:\ast \xi_{\lambda\tau}\:\partial^{\tau}\:(S)\:S^{-1}] \right) \ .
\label{S2L}
\end{eqnarray}

We would like to simplify the notation by calling,

\begin{eqnarray}
S_{(1)}^{'\mu} &=& \epsilon^{\mu\nu}\:\epsilon^{\sigma}_{\:\:\nu}\: Tr[\tilde{\Lambda}^{\alpha}_{\:\:\:\delta}\:\tilde{\Lambda}^{\beta}_{\:\:\:\gamma}\:\Sigma^{\delta\gamma}\:E_{\alpha}^{\:\:\rho}\: E_{\beta}^{\:\:\lambda}\:\ast \xi_{\rho\sigma}\:\ast \xi_{\lambda\tau}\:A^{\tau}] \label{S1TILDE}\\
S_{(2)}^{'\mu} &=& \sqrt{-Q_{ym}/2} \: \epsilon^{\mu\sigma} \: Tr[\tilde{\Lambda}^{\alpha}_{\:\:\:\delta}\:\tilde{\Lambda}^{\beta}_{\:\:\:\gamma}\:\Sigma^{\delta\gamma}\:E_{\alpha}^{\:\:\rho}\: E_{\beta}^{\:\:\lambda}\:\ast \xi_{\rho\sigma}\:\ast \xi_{\lambda\tau}\:A^{\tau}] \ . \label{S2TILDE}
\end{eqnarray}

Then, we can write equations (\ref{S1L}-\ref{S2L}) as,

\begin{eqnarray}
\tilde{S}_{(1)}^{\mu} &=& S_{(1)}^{'\mu} + C^{'}\: S_{(1)}^{'\mu} + D^{'}\:S_{(2)}^{'\mu}
\label{S1SHORT}\\
\tilde{S}_{(2)}^{\mu} &=& S_{(2)}^{'\mu} + E^{'}\: S_{(1)}^{'\mu} + F^{'}\:S_{(2)}^{'\mu}\ . \label{S2SHORT}
\end{eqnarray}

We could have written $\tilde{S}_{(1)}^{\mu}$ and $\tilde{S}_{(2)}^{\mu}$ in terms of $S_{(1)}^{\mu}$ and $S_{(2)}^{\mu}$. Instead we wrote them in terms of $S_{(1)}^{'\mu}$ and $S_{(2)}^{'\mu}$ because it is more convenient and clear for our subsequent steps in this particular section.
Again, if we carefully follow the steps in the section gauge geometry in \cite{A}, we can conclude that,

\begin{eqnarray}
E^{'} &=& D^{'} \\
F^{'} &=& C^{'} \ ,
\end{eqnarray}

where,

\begin{eqnarray}
C^{'} &=&{\imath \over g}\: (-Q_{ym}/2)\:\:S^{'\sigma\:}_{(1)}\: Tr[\tilde{\Lambda}^{\alpha}_{\:\:\:\delta}\:\tilde{\Lambda}^{\beta}_{\:\:\:\gamma}\:\Sigma^{\delta\gamma}\:E_{\alpha}^{\:\:\rho}\: E_{\beta}^{\:\:\lambda}\:\ast \xi_{\rho\sigma}\:\ast \xi_{\lambda\tau}\:\partial^{\tau}\:(S)\:S^{-1}] / (\:S^{'}_{(2)\mu}\:S_{(2)}^{'\mu}\:) \label{COEFFCPRIMA}\\
D^{'} &=&{\imath \over g}\: (-Q_{ym}/2)\:\:S^{'\sigma\:}_{(2)}\: Tr[\tilde{\Lambda}^{\alpha}_{\:\:\:\delta}\:\tilde{\Lambda}^{\beta}_{\:\:\:\gamma}\:\Sigma^{\delta\gamma}\:E_{\alpha}^{\:\:\rho}\: E_{\beta}^{\:\:\lambda}\:\ast \xi_{\rho\sigma}\:\ast \xi_{\lambda\tau}\:\partial^{\tau}\:(S)\:S^{-1}] / (\:S^{'}_{(1)\mu}\:S_{(1)}^{'\mu}\:) \ . \label{COEFFDPRIMA}
\end{eqnarray}

We would like as well, to calculate the norm of the transformed vectors
$\tilde{S}_{(1)}^{\mu}$ and $\tilde{S}_{(2)}^{\mu}$,

\begin{eqnarray}
\tilde{S}_{(1)}^{\mu}\:\tilde{S}_{(1)\mu} &=&
[(1+C^{'})^{2}-D^{'2}]\:S_{(1)}^{'\mu}\:S^{'}_{(1)\mu}\label{FP}\\
\tilde{S}_{(2)}^{\mu}\:\tilde{S}_{(2)\mu} &=&
[(1+C^{'})^{2}-D^{'2}]\:S_{(2)}^{'\mu}\:S^{'}_{(2)\mu}\ ,\label{SP}
\end{eqnarray}

where the relation $S_{(1)}^{'\mu}\:S^{'}_{(1)\mu}=
-S_{(2)}^{'\mu}\:S^{'}_{(2)\mu}$ has been used.

It is possible at this point to repeat all the discussion about the different cases that arise according to the sign of $(1+C^{'})^{2}-D^{'2}$, as it was done in \cite{A}. Regarding equations (\ref{FP}-\ref{SP}), there is in the first paper ``Tetrads in geometrodynamics'' \cite{A} a full discussion concerning the factor on the right hand side (section Gauge transformations on blade one). It would be redundant to repeat it here. It is also straightforward to understand that a similar analysis can be done on blade two, for the transformation of the vectors $S_{(3)}^{\mu}$ and $S_{(4)}^{\mu}$ that we are assumming to be spacelike. Our first conclusion from the results above, is that $SU(2)$ local gauge transformations, generate the composition of several transformations. First, there is a local tetrad transformation, generated by a locally inertial coordinate transformation $\tilde{\Lambda}^{\alpha}_{\:\:\:\delta}$, of the electromagnetic tetrads $E_{\alpha}^{\rho}$. Second, the normalized tetrad vectors $S_{(1)}^{'\mu}$ and $S_{(2)}^{'\mu}$, undergo a LB1 transformation on the blade they generate. The two normalized vectors $\tilde{S}_{(1)}^{\mu}$ and $\tilde{S}_{(2)}^{\mu}$, end up on the same blade one, generated by the original normalized generators of the blade, $\left({S_{(1)}^{\mu} \over \sqrt{-S_{(1)}^{\nu}\:S_{(1)\nu}}}, {S_{(2)}^{\mu} \over \sqrt{S_{(2)}^{\nu}\:S_{(2)\nu}}}\right)$, as it was highlighted at the beginning of this section. Therefore, the gauge invariance of the metric tensor is assured. We can continue making several important remarks about these tetrad transformations. Within the set of LB1 tetrad transformations of the $\left({S_{(1)}^{\mu} \over \sqrt{-S_{(1)}^{\nu}\:S_{(1)\nu}}}, {S_{(2)}^{\mu} \over \sqrt{S_{(2)}^{\nu}\:S_{(2)\nu}}}\right)$, there is an identity transformation that corresponds to the identity in $SU(2)$. To every LB1 tetrad transformation, which in turn is generated by $S$ in $SU(2)$, there corresponds an inverse, generated by $S^{-1}$. We observe also the following. Since locally inertial coordinate transformations $\tilde{\Lambda}^{\alpha}_{\:\:\:\delta}$ of the electromagnetic tetrads $E_{\alpha}^{\rho}$ in general do not commute, then the locally $SU(2)$ electromagnetic tetrad generated transformations are non-Abelian. The non-Abelianity of $SU(2)$ is mirrored by the non-commutativity of these locally inertial coordinate transformations $\tilde{\Lambda}^{\alpha}_{\:\:\:\delta}$. The key role in this non-commutativity is played by the object $\Sigma^{\alpha\beta}$, that translates local $SU(2)$ gauge transformations, into locally inertial Lorentz transformations. Another issue of relevance is related to the analysis of the ``memory'' of these transformations. We would like to know explicitly, if a second LB1 tetrad transformation, generated by a local gauge transformation $S_{2}$, is going to ``remember'' the existence of the first one, generated by $S_{1}$. To this end, let us just write for instance the vector $\tilde{\tilde{S}}_{(1)}^{\:\mu}$ after these two gauge transformations,

\begin{eqnarray}
\tilde{\tilde{S}}_{(1)}^{\:\mu} = \epsilon^{\mu\nu}\:\epsilon^{\sigma}_{\:\:\nu}\: Tr[\tilde{\Lambda}^{\alpha}_{2\:\:\kappa}\:\tilde{\Lambda}^{\beta}_{2\:\:\epsilon}\:\tilde{\Lambda}^{\kappa}_{1\:\:\delta}\:\tilde{\Lambda}^{\epsilon}_{1\:\:\gamma}\:\Sigma^{\delta\gamma}\:E_{\alpha}^{\:\:\rho}\: E_{\beta}^{\:\:\lambda}\:\ast \xi_{\rho\sigma}\:\ast \xi_{\lambda\tau}\:A^{\tau}] + \nonumber \\
{\imath \over g} \:\epsilon^{\mu\nu}\:\epsilon^{\sigma}_{\:\:\nu}\: Tr[\tilde{\Lambda}^{\alpha}_{2\:\:\kappa}\:\tilde{\Lambda}^{\beta}_{2\:\:\epsilon}\:\tilde{\Lambda}^{\kappa}_{1\:\:\delta}\:\tilde{\Lambda}^{\epsilon}_{1\:\:\gamma}\:\Sigma^{\delta\gamma}\:E_{\alpha}^{\:\:\rho}\: E_{\beta}^{\:\:\lambda}\:\ast \xi_{\rho\sigma}\:\ast \xi_{\lambda\tau}\:\partial^{\tau}\:(S_{1})\:S_{1}^{-1}] + \nonumber \\
{\imath \over g} \:\epsilon^{\mu\nu}\:\epsilon^{\sigma}_{\:\:\nu}\: Tr[\tilde{\Lambda}^{\alpha}_{2\:\:\delta}\:\tilde{\Lambda}^{\beta}_{2\:\:\gamma}\:\Sigma^{\delta\gamma}\:E_{\alpha}^{\:\:\rho}\: E_{\beta}^{\:\:\lambda}\:\ast \xi_{\rho\sigma}\:\ast \xi_{\lambda\tau}\:\partial^{\tau}\:(S_{2})\:S_{2}^{-1}] \ .
\label{S1DOUBLETR}
\end{eqnarray}

We can notice that the second term contains the same electromagnetic transformed tetrads as the first one. Therefore, when we compare these two terms, it is straightforward to see that it is not possible, after the second gauge transformation, from these transformed electromagnetic tetrads, to ``remember'' any relative change associated to the second gauge transformation. In addition, the second line in (\ref{S1DOUBLETR}) contains only $S_{1}$, and the third line contains only $S_{2}$. This means that the second LB1 tetrad transformation on blade one is not going to remember the first one. The algebra underlying these statements can be followed through \cite{A}. Then, another way of thinking of (\ref{S1DOUBLETR}) is by first performing two successive local Lorentz transformations of the electromagnetic tetrad in the first line, and second, by performing two successive $LB1$ tetrad transformations in the second and third line. In section \ref{sec:appIII} we study one last important remark about the composition of transformations. Let us analyze the expression $\tilde{E}_{\delta}^{\:\:\rho} = \tilde{\Lambda}^{\alpha}_{\:\:\delta}\:E_{\alpha}^{\:\:\rho}$. This is going to be a Lorentz transformed electromagnetic tetrad vector. The Lorentz transformations are generated by $SU(2)$ local gauge transformations, and can be thought of, as simple spatial rotations, see section \ref{sec:appII}.
Then, if we keep the same notation as in \cite{A}, we can call,

\begin{eqnarray}
\tilde{\xi}^{\mu\nu} &=& -2\:\sqrt{-Q/2}\:\tilde{\Lambda}^{\delta}_{\:\:o}\:\tilde{\Lambda}^{\gamma}_{\:\:1}\:E_{[\delta}^{\:\:\mu}\:E_{\gamma]}^{\:\:\nu}\label{ET}\\
\ast \tilde{\xi}^{\mu\nu} &=& 2\:\sqrt{-Q/2}\:\tilde{\Lambda}^{\delta}_{\:\:2}\:\tilde{\Lambda}^{\gamma}_{\:\:3}\:E_{[\delta}^{\:\:\mu}\:E_{\gamma]}^{\:\:\nu}\ .\label{DET}
\end{eqnarray}

Now, with these fields, the $\tilde{\xi}_{\mu\nu}$, and its dual $\ast \tilde{\xi}_{\mu\nu}$, we can repeat the procedure followed in \cite{A}, and the transformed tetrads $\tilde{E}_{\alpha}^{\:\:\rho}$, can be rewritten completely in terms of these ``new'' extremal fields. It is straightforward to prove that $\tilde{\xi}^{\mu\lambda}\:\ast \tilde{\xi}_{\mu\nu} = 0$.
It is also evident that $\tilde{E}_{o}^{\:\:\mu} \:\ast \tilde{\xi}_{\mu\nu} = 0 = \tilde{E}_{1}^{\:\:\mu} \:\ast \tilde{\xi}_{\mu\nu}$. Therefore $\tilde{E}_{o}^{\:\:\mu}$ and $\tilde{E}_{1}^{\:\:\mu}$ belong to the plane generated by the normalized version of vectors like $\tilde{\xi}^{\mu\nu}\:\tilde{\xi}_{\lambda\nu}\:X^{\lambda}$ and $\tilde{\xi}^{\mu\nu}\:X_{\nu}$.
Then, for instance we are going to be able to write the timelike $\tilde{E}_{o}^{\:\:\mu}$ as the the normalized version of the timelike $\tilde{\xi}^{\mu\nu}\:\tilde{\xi}_{\lambda\nu}\:\tilde{X}^{\lambda}$ for some vector field  $\tilde{X}^{\lambda}$. We remind ourselves that the relation between the normalized versions of the two vectors that locally determine blade one, $\tilde{\xi}^{\mu\nu}\:\tilde{\xi}_{\lambda\nu}\:X^{\lambda}$ and  $\tilde{\xi}^{\mu\nu}\:X_{\nu}$ on one hand, and $\tilde{\xi}^{\mu\nu}\:\tilde{\xi}_{\lambda\nu}\:\tilde{X}^{\lambda}$ on the other hand, is established through a LB1 gauge transformation \cite{A}. Analogous analysis for $\tilde{E}_{2}^{\:\:\mu}$ and $\tilde{E}_{3}^{\:\:\mu}$ on blade two.  Gauge transformations of the electromagnetic tetrads we remind ourselves are nothing but a special kind of tetrad transformations that belong either to the groups LB1 or LB2.

We know that there are three sets of tetrads in correspondence to the existence of three generators in the $SU(2)$ gauge group. These tetrads could be gauge dependent in the sense of section \ref{3complexions}, or gauge dependent in the sense of section \ref{extremalsu2}. The gauge transformation properties of the tetrads introduced in section \ref{extremalsu2} are convenient in the sense that they are going to transform under $SU(2)$ gauge transformations following the same geometrically transparent pattern of the Abelian tetrads introduced in a spacetime where only an electromagnetic field is present \cite{A}. Once we make three choices for the tensor $\omega_{\mu\nu}$, out of all the possible ones, we can study the mutual relation between the $SU(2)$ group, and the three LB1 (or LB2) groups, associated to our already chosen three sets of tetrads. Let us then study the LB1 transformations for one of these sets of tetrads. For all the other tetrads and also for the LB2 rotations the analysis is just analogous.
For each $SU(2)$ element $S$, there exist three local scalar functions $\theta^{i}$, $i=1\ldots3$, see section \ref{sec:appI}. Borrowing the notation and line of thinking from the section group isomorphism in \cite{A}, we can write a set of three equations relating these three $SU(2)$ local functions and the corresponding three LB1 scalar functions $\phi^{'\,(h)}$, $h=1\ldots3$.

\begin{eqnarray}
{\imath \over g}\: (-Q_{ym}^{(h)} / 2)\:Tr[\Sigma^{\alpha\beta}\:E_{\alpha}^{\:\:\rho}\: E_{\beta}^{\:\:\lambda}\:\ast \xi_{\rho\sigma}\:\ast \xi_{\lambda\tau}\:S^{-1}\:\partial^{\tau}\:(S)] \nonumber \\ = {\imath \over g}\:
(-Q_{ym}^{(h)} / 2)\:Tr[\tilde{\Lambda}^{\alpha}_{\:\:\:\delta}\:\tilde{\Lambda}^{\beta}_{\:\:\:\gamma}\:\Sigma^{\delta\gamma}\:E_{\alpha}^{\:\:\rho}\: E_{\beta}^{\:\:\lambda}\:\ast \xi_{\rho\sigma}\:\ast \xi_{\lambda\tau}\:\partial^{\tau}\:(S)\:S^{-1}] \nonumber \\ = -C^{'\,(h)}\:S_{(1)\sigma}^{'\,(h)} - D^{'\,(h)}\: S_{(2)\sigma}^{'\,(h)} + M^{'\,(h)}\: S_{(3)\sigma}^{'\,(h)} + N^{'\,(h)}\: S_{(4)\sigma}^{'\,(h)}\ , \label{groupmorph}
\end{eqnarray}

such that,

\begin{eqnarray}
D^{'\,(h)} &=& (1+C^{'\,(h)})\: \tanh\phi^{'\,(h)} \;\;\;\mbox{for proper transformations}\label{DP}\\
D^{'\,(h)} &=& (1+C^{'\,(h)}) / \tanh\phi^{'\,(h)} \;\;\;\mbox{for improper transformations}\ . \label{DI}
\end{eqnarray}

The index $h$ runs from one to three, representing the three sets of tetrads, and the summation convention is not applied on $(h)$. $M^{'\,(h)}$ and $N^{'\,(h)}$ arise for blade two in a similar fashion as $C^{'\,(h)}$ and $D^{'\,(h)}$ arise for blade one \cite{A}. Therefore, (\ref{groupmorph}) are three sets of equations that relate the three local $\theta^{i}$, implicitly included in $S$, and the three local $\phi^{'\,(h)}$.
We would like to analyze one more issue. Let us suppose that we map the local gauge group $SU(2)$ into the three LB1 groups. We are interested in the injectivity of such a group mapping. Let us suppose then, that $S_{1}$ and $S_{2}$ generate the same $\phi^{'\,(h)}$
for $h=1\ldots3$. The product $S_{1}\:S_{2}^{-1}$ should generate the identity, meaning that $S_{1} = S_{2}$. Therefore, the injectivity of the mapping remains proved. In section \ref{sec:appI} it is also proved that the image of this group mapping is not a subgroup of the three LB1 groups. Thus, we are able now to formulate the following results,

\newtheorem {guesslb1} {Theorem}
\newtheorem {guesslb2}[guesslb1] {Theorem}
\begin{guesslb1}
The mapping between the local gauge group $SU(2)$ of transformations and the tensor product of the three local groups of LB1 tetrad transformations is isomorphic. \end{guesslb1}

Following analogously the reasoning laid out in \cite{A}, in addition to the ideas above, we can also state,

\begin{guesslb2}
The mapping between the local gauge group $SU(2)$ of transformations and the tensor product of the three local groups of LB2 tetrad transformations is isomorphic. \end{guesslb2}

\section{Applications}
\label{appli}

\subsection{Gauge invariants}
\label{Gauge invariants}

First of all we would like to introduce new gauge invariant objects built out of the tetrad components of the stress-energy tensor. Only in this section when we write $T_{\mu\nu}$ we mean $T^{(ym)}_{\mu\nu}$, we just do not want to overload the equations with notation. Given the tetrad $W_{(o)}^{\mu}$, $W_{(1)}^{\mu}$, $W_{(2)}^{\mu}$, $W_{(3)}^{\mu}$, (no confusion should arise with vector $E_{3}^{\:\:\rho} = W^{\rho}$ which is just one vector in the electromagnetic tetrad) which we consider to be the normalized version of $S_{(1)}^{\mu}$, $S_{(2)}^{\mu}$, $S_{(3)}^{\mu}$, $S_{(4)}^{\mu}$, we perform the gauge transformations on blades one and two,

\begin{eqnarray}
\tilde{W}_{(o)}^{\mu} &=& \cosh\phi\:W_{(o)}^{\mu} + \sinh\phi\:W_{(1)}^{\mu}\label{GT1}\\
\tilde{W}_{(1)}^{\mu} &=& \sinh\phi\:W_{(o)}^{\mu} + \cosh\phi\:W_{(1)}^{\mu}\label{GT2}\\
\tilde{W}_{(2)}^{\mu} &=& \cos\psi\:W_{(2)}^{\mu} - \sin\psi\:W_{(3)}^{\mu}\label{GT3}\\
\tilde{W}_{(3)}^{\mu} &=& \sin\psi\:W_{(2)}^{\mu} + \cos\psi\:W_{(3)}^{\mu}\ .\label{GT4}
\end{eqnarray}

It is a matter of algebra to prove that the following objects are invariant under the set of transformations (\ref{GT1}-\ref{GT4}),

\begin{eqnarray}
\lefteqn{ \left(\:W_{(0)}^{\mu}\:T_{\mu\nu}\:W_{(0)}^{\nu}\right)\:W_{(0)}^{\lambda}\:W_{(0)}^{\rho} - \left(\:W_{(0)}^{\mu}\:T_{\mu\nu}\:W_{(1)}^{\nu}\right)\:\left[W_{(0)}^{\lambda}\:W_{(1)}^{\rho} + W_{(0)}^{\rho}\:W_{(1)}^{\lambda}\right] + } \nonumber \\
&&\left(\:W_{(1)}^{\mu}\:T_{\mu\nu}\:W_{(1)}^{\nu}\right)\:W_{(1)}^{\lambda}\:W_{(1)}^{\rho}  \label{GI1} \\
&&-\left(\:W_{(0)}^{\mu}\:T_{\mu\nu}\:W_{(2)}^{\nu}\right)\:\left[W_{(0)}^{\lambda}\:W_{(2)}^{\rho} + W_{(0)}^{\rho}\:W_{(2)}^{\lambda}\right]
-\left(\:W_{(0)}^{\mu}\:T_{\mu\nu}\:W_{(3)}^{\nu}\right)\:\left[W_{(0)}^{\lambda}\:W_{(3)}^{\rho} + W_{(0)}^{\rho}\:W_{(3)}^{\lambda}\right] + \nonumber \\
&&\left(\:W_{(1)}^{\mu}\:T_{\mu\nu}\:W_{(2)}^{\nu}\right)\:\left[W_{(1)}^{\lambda}\:W_{(2)}^{\rho} + W_{(1)}^{\rho}\:W_{(2)}^{\lambda}\right] + \
\left(\:W_{(1)}^{\mu}\:T_{\mu\nu}\:W_{(3)}^{\nu}\right)\:\left[W_{(1)}^{\lambda}\:W_{(3)}^{\rho} + W_{(1)}^{\rho}\:W_{(3)}^{\lambda}\right] \label{GI2}\\
&&\left(\:W_{(2)}^{\mu}\:T_{\mu\nu}\:W_{(2)}^{\nu}\right)\:W_{(2)}^{\lambda}\:W_{(2)}^{\rho} + \left(\:W_{(2)}^{\mu}\:T_{\mu\nu}\:W_{(3)}^{\nu}\right)\:\left[W_{(2)}^{\lambda}\:W_{(3)}^{\rho} + W_{(2)}^{\rho}\:W_{(3)}^{\lambda}\right] + \nonumber \\
&&\left(\:W_{(3)}^{\mu}\:T_{\mu\nu}\:W_{(3)}^{\nu}\right)\:W_{(3)}^{\lambda}\:W_{(3)}^{\rho}\ .\label{GI3}
\end{eqnarray}

The subtlety here is the following. Using any normalized tetrads, and under tetrad transformations of the kind (\ref{GT1}-\ref{GT4}), the objects (\ref{GI1}-\ref{GI3}) are going to remain invariant. The point is that the transformations (\ref{GT1}-\ref{GT4}), are tetrad gauge transformations, or tetrad gauge generated trasformations, see section \ref{gaugegeometry} and \cite{A}. It is the way in which the normalized version of tetrad vectors (\ref{S1}-\ref{S4}) transform on blades one and two under locally generated $SU(2)$ gauge transformations. The tensor $T_{\mu\nu}$ is gauge invariant by itself as we already know. Then these are true new gauge invariants under (\ref{GT1}-\ref{GT4}). We might wonder what happens with the objects (\ref{GI1}-\ref{GI3}), when we perform discrete gauge transformations on blade one. It is evident that all of the objects remain invariant under a tetrad full inversion on blade one. However, under the discrete transformation represented by equations (64-65) in reference \cite{A}, while objects (\ref{GI1}) and (\ref{GI3}) remain invariant, object (\ref{GI2}) changes in a global sign (gets multiplied globally by $-1$). Therefore we can say that objects (\ref{GI1}) and (\ref{GI3}) are true and new gauge invariants, while object (\ref{GI2}) is invariant under boosts generated gauge transformations on blade one, rotations on blade two, full inversions on blade one, but gets multiplied by $-1$ under the discrete gauge generated transformation on blade one given by equations (64-65) in reference \cite{A}. We are going to make use of these gauge invariant properties of objects (\ref{GI1}-\ref{GI3}) in the next section that deals with the diagonalization of the stress-energy tensor.

\subsection{Diagonalization of the stress-energy tensor}
\label{diagonal}

We proceed now to extend to the non-Abelian case the algorithm for the diagonalization of the stress-energy tensor. The Abelian case was studied thoroughly in manuscript \cite{A}. It is worth mentioning that we are going to present one method, but there are others, all equivalent of course. Another method that we call multiple extremal representation of the gravitational field will be introduced in an upcoming paper. In the previous section \ref{Gauge invariants} we found that we can build with the stress-energy tensor and the new tetrads, three objects that are locally gauge invariant. This is a mathematical truth that can be easily checked. Then, we might ask about the usefulness of the existence of these three new gauge invariant objects, and our answer is the following. These three new local gauge invariant objects allow us to connect gauge invariance with three different blocks in the stress-energy tensor. One block off-diagonal and two diagonal blocks, separately. Therefore these three new gauge invariant objects are going to guide us in establishing a local gauge invariant process of diagonalization of the stress-energy tensor. Their existence means that we can block diagonalize the stress-energy tensor in a gauge invariant way, locally. We start by putting forward a generalized duality transformation for non-Abelian fields. For instance we might choose,

\begin{equation}
\varepsilon_{\mu\nu} =  Tr[\vec{n}\: \cdot \: f_{\mu\nu} - \vec{l}\: \cdot \: \ast f_{\mu\nu}] \ ,\label{gendual}
\end{equation}

where $f_{\mu\nu} = f^{a}_{\mu\nu}\:\sigma^{a}$, $\vec{n} = n^{a}\:\sigma^{a}$ and $\vec{l} = l^{a}\:\sigma^{a}$ are vectors in isospace. The $\cdot$ means product in isospace. $\sigma^{a}$ are the pauli matrices see \ref{sec:appI} and the summation convention is applied on the internal index $a$. The vector components are defined as,

\begin{eqnarray}
\lefteqn{ \vec{n} = (\cos\theta_{1},\cos\theta_{2},\cos\theta_{3}) } \label{ISO1} \\
&&\vec{l} = (\cos\beta_{1},\cos\beta_{2},\cos\beta_{3}) \ , \label{ISO2}
\end{eqnarray}

where all the six isoangles are local scalars that satisfy,

\begin{eqnarray}
\lefteqn{ \Sigma_{a=1}^{3} \cos^{2}\theta_{a} = 1 } \label{ISOSUM1} \\
&&\Sigma_{a=1}^{3} \cos^{2}\beta_{a} = 1 \ . \label{ISOSUM2}
\end{eqnarray}

In isospace $\vec{n} = n^{a}\:\sigma^{a}$ transforms under a local $SU(2)$ gauge transformation $S$, as $S^{-1}\:\vec{n}\:S$, see chapter III in \cite{CBDW} and also reference \cite{GRSYMM}, and similar for $\vec{l} = l^{a}\:\sigma^{a}$. The tensor $f_{\mu\nu} = f^{a}_{\mu\nu}\:\sigma^{a}$ transforms as
$f_{\mu\nu} \rightarrow S^{-1}\:f_{\mu\nu}\:S$. Therefore $\varepsilon_{\mu\nu}$ is manifestly gauge invariant. We can see from (\ref{ISO1}-\ref{ISO2}) and (\ref{ISOSUM1}-\ref{ISOSUM2}) that only four of the six angles in isospace are independent. Next we perform one more duality transformation,

\begin{equation}
\Omega_{\mu\nu} = \cos\alpha_{d} \:\: \varepsilon_{\mu\nu} -
\sin\alpha_{d} \:\: \ast \varepsilon_{\mu\nu} \ ,\label{diagdual}
\end{equation}

such that the complexion $\alpha_{d}$ is defined by the usual local condition $\Omega_{\mu\nu}\:\ast \Omega^{\mu\nu} = 0$, see reference \cite{A},

\begin{eqnarray}
\tan(2\alpha_{d}) = - \varepsilon_{\mu\nu}\:\ast \varepsilon^{\mu\nu} / \varepsilon_{\lambda\rho}\:\varepsilon^{\lambda\rho}\ .\label{compdd}
\end{eqnarray}

All the conclusions derived in \cite{A} are valid in this context and therefore exactly as in reference \cite{A}. Using the local antisymmetric tensor $\Omega_{\mu\nu}$, we can produce tetrad skeletons and with new gauge vectors $X_{d}^{\sigma}$ and $Y_{d}^{\sigma}$ we can build a new normalized tetrad. This new tetrad that we call $T_{\alpha}^{\mu}$ has four independent isoangles included in its definition, in the skeletons. There is also the freedom to introduce an LB1 and an LB2 local  $SU(2)$ generated transformations on both blades by new angles $\phi_{d}$ and $\psi_{d}$ (through the gauge vectors $X_{d}^{\sigma}$ and $Y_{d}^{\sigma}$) which are not yet fixed and represent two more independent angles. Having six independent and undefined angles, we are going to use this freedom to choose them when fixing the six diagonalization conditions for the stress-energy tensor. It must be highlighted and stressed that since the local antisymmetric tensor $\Omega_{\mu\nu}$ is gauge invariant, then the tetrad vectors skeletons are $SU(2)$ gauge invariant. This was a fundamental condition that we made in previous sections in order to ensure the metric invariance when performing LB1 and LB2 transformations. Then, we proceed to impose the diagonalization conditions,

\begin{eqnarray}
\lefteqn{ T_{o1} = T_{o}^{\mu}\:T_{\mu\nu}\: T_{1}^{\nu} = 0 } \label{diagcond1} \\
&&T_{o2} = T_{o}^{\mu}\:T_{\mu\nu}\: T_{2}^{\nu} = 0 \label{diagcond2} \\
&&T_{o3} = T_{o}^{\mu}\:T_{\mu\nu}\: T_{3}^{\nu} = 0 \label{diagcond3} \\
&&T_{12} = T_{1}^{\mu}\:T_{\mu\nu}\: T_{2}^{\nu} = 0 \label{diagcond4} \\
&&T_{13} = T_{1}^{\mu}\:T_{\mu\nu}\: T_{3}^{\nu} = 0 \label{diagcond5} \\
&&T_{23} = T_{2}^{\mu}\:T_{\mu\nu}\: T_{3}^{\nu} = 0 \ .\label{diagcond6}
\end{eqnarray}

These are finally the six equations that locally define the six angles $\theta_{1},\:\theta_{2},\:\beta_{1},\:\beta_{2},\:\phi_{d},\:\psi_{d}$, for instance. The other two $\theta_{3},\:\beta_{3}$ are determined by equations (\ref{ISOSUM1}-\ref{ISOSUM2}) once the other six have already been determined through equations (\ref{diagcond1}-\ref{diagcond6}). Once the stress-energy tensor has been diagonalized, always assuming that the local diagonalization process is possible, in the new gauge, the ``diagonal gauge'', we can study the gauge invariants (\ref{GI1}-\ref{GI3}). We imposed the off-diagonal tetrad components of the stress-energy tensor (\ref{diagcond2}-\ref{diagcond5}) to be zero. These four equations are manifestly and locally $SU(2)$ gauge invariant by themselves under LB1 and LB2 local transformations of the vectors $T_{\alpha}^{\mu}$, analogous to transformations (\ref{GT1}-\ref{GT4}). Therefore the new off-diagonal gauge invariant object (\ref{GI2}), built with the stress-energy tensor off-diagonal tetrad components, is also zero locally. It is consistent because this object is precisely invariant under $SU(2)$ local gauge transformations (except for a global negative sign in one particular discrete case, which makes no difference if the object is zero). Then, we conclude, if its components are all null, zero in one gauge, in this case the ``diagonal gauge'', they all will be null in any other gauge. The two remaining blocks associated to the two remaining locally gauge invariant objects in the diagonal of the stress-energy tensor, are next diagonalized by suitable tetrad rotations in the planes one and two through the use of the gauge vectors $X_{d}^{\sigma}$ and $Y_{d}^{\sigma}$. That is, by $SU(2)$ tetrad gauge transformations on these planes, that have been proven to be equivalent to tetrad Lorentz transformations LB1 and LB2 on these planes. This is done by imposing conditions (\ref{diagcond1}) and (\ref{diagcond6}). See \cite{A} and the section gauge geometry in this present manuscript. The other two objects (\ref{GI1}) and (\ref{GI3}) will be maximally simplified since the off-diagonal terms in both of them will vanish in the ``diagonal gauge''. It is evident that the ``diagonal gauge'' might be a source of simplification in dealing with the field equations, and of course the inherent simplification in the geometrical analysis of any problem involving these kind of fields (\ref{eyme}-\ref{ymvfe2}).

\section{Conclusions}

The $SU(2)$ local gauge group of transformations associated with Yang-Mills fields, finds its counterpart in geometrical structures. To find this relation between gauge, and geometrical structures we build in a succession, different sets of tetrad vector fields. First, the three extremal fields and complexions that arise from the diagonalization of each component of the Yang-Mills stress-energy tensor allow for the construction of three sets of tetrad vectors that have a similar structure than their Abelian counterparts \cite{A}, but lack the key property they have. This property is related to the fact that in the Abelian environment associated with electromagnetic fields, the local gauge transformation of the tetrad vectors induces a LB1 Lorentz transformation on blade one, such that the two vectors that generate this blade, remain on the blade after the transformation. Similar for rotations on blade two. This property is essential as far as we ask for the metric tensor to remain invariant under $U(1)$ transformations in the Abelian case. We demand a similar property for the metric tensor in spacetimes where $SU(2)$ Yang-Mills fields are present. That is the reason why we take on the task of finding tetrads that have transformation properties analogous to the Abelian, in this non-Abelian environment. Once we build these new tetrads in section \ref{extremalsu2}, we study their transformation properties. They have an inherent freedom in the choice of two vector fields. These two vector fields are available freedom in the construction of our tetrads. They are ``gauge'' by themselves, and they include ``gauge'' in their construction. It is this freedom we are exploiting in order to prove our results, and it is this freedom the one that allow us to see geometrically in a transparent way how we can translate the abstract internal local group of transformations into spacetime local groups of transformations. These vectors chosen for this particular example in $SU(2) \times U(1)$ Yang-Mills geometrodynamics, clearly show in a few steps, that there is a group morphism between $SU(2)$, and LB1. Analogous for LB2. In fact there is a morphism for each tetrad, and we learnt in section \ref{3complexions} that it is possible to consider as many tetrads, as generators has the gauge group. These group mappings clearly show the relation between the local gauge structures and their geometrical counterparts.

The physical and geometrical significance of this work reside in the following issues.

\begin{enumerate}
\item It was not known before, an explicit relationship between gauge and gravity. It was known the relationship between $SU(2)$ and $SO(3)$, but not the explicit relationship between $SU(2)$ and the gravitational field. As in the first paper in the tetrad series, the fundamental issue is placed in this relationship. The field equations (\ref{eyme}-\ref{ymvfe2}) above stated are not just by themselves hinting us about the possibility of constructing the kind of tetrads that we are presenting in our works. These tetrads allow for an explicit and direct way to see how the spacetime geometry relates to the ``internal'' local gauge groups of transformations, so far associated to microparticle structures through the standard model \cite{DG}$^{,}$\cite{GR}, but not to any kind of spacetime local groups of transformations that explicitly exhibit the invariance of the gravitational field. Therefore we are introducing for the first time and explicit ``link'' between the ``internal'' and the ``spacetime'', so far detached from each other. Moreover, since the gravitational field is invariant under the local groups of spacetime transformations LB1 and LB2, which in turn are generated by the internal local group of transformations $SU(2)$, one may wonder if particle multiplets can be associated to gravitational fields which are explicitly invariant under these groups of local transformations \cite{GRSYMM}. %We may comment for instance on a geometry electromagnetically neutral where we can always choose the vectors $X^{\mu} = Tr[f^{\mu\lambda}\:A_{\lambda}] %= f^{k\mu\lambda}\:\ast A^{k}_{\lambda}$, and $Y^{\mu} = Tr[f^{\mu\lambda}\:\ast A_{\lambda}] = f^{k\mu\lambda}\:\ast A^{k}_{\lambda}$, where the %summation convention was applied on the internal index $k$.
    The microparticles would then be tetrad gauge ``states'' of the gravitational fields. The needles pointing in internal abstract spaces into different microparticle states, would be translated in our tetrad language, into tetrad needles pointing locally into different directions, which are related to each other through local ``rotations'' LB1, LB2, which in turn are generated by internal local gauge transformations.

\item In this manuscript we are settling the issue about the relation between the groups so far regarded as generating local ``internal'' transformations, and local ``spacetime'' transformations. This is not a minor issue. For over eighty years the ``internal'' has been regarded as detached from the ``spacetime''. The standard model has been designed on the pillar of gauge invariance. Finding this relation amounts to finding the relationship to the gravitational field. The local tetrads will provide the metric tensor and so on. But the relevant point is that now we know that the ``link'' between internal structures and spacetime structures is bridged by tetrads, and local gauge transformations are isomorphic to local tetrad transformations that explicitly leave the metric tensor invariant. Think then of a new problem involving spinors and a new choice for the two gauge vectors $Y^{\mu} = X^{\mu} = E_{\alpha}^{\mu}\:{\overline \psi}\:\sigma^{\alpha}\:\psi$. We can even think of this choice in many problems involving spinors $\psi$. A gauge transformation of the spinors would be translated into a locally inertial transformation of the electromagnetic tetrad $ E_{\alpha}^{\mu}$, for instance. $Y^{\mu} = X^{\mu} \rightarrow \Lambda_{\:\:\:\beta}^{\alpha}\:E_{\alpha}^{\mu}\:{\overline \psi}\:\sigma^{\beta}\:\psi$. In turn this would directly imply that when the states represented by the spinor field, locally transform, the tetrads rotate in blades one and two. This is then, a direct and explicit ``link'' between the spinor states and the states of the gravitational field.

%\item There is also the issue of the Higgs mechanism \cite{DG}\cite{GR}. This issue is explored in another work ``Tetrads in low-energy weak interactions'' \cite{A6} . The point is that we are able to associate to an interacting process like Inverse Muon decay or elastic Neutrino-Electron scattering, a gravitational field. Then, one may wonder if the Higgs mechanism is truly necessary because the mass of the mass mediators could just be associated to the gravitational field associated to the interacting process. We strongly believe that these physical possibilities are justifying the tetrad formalism presented in this series of papers.

\item It has been thought for a long time  \cite{SWNG}$^{,}$\cite{LORNG}$^{,}$\cite{CMNG}$^{,}$\cite{MK} that there is no relationship between the spacetime and internal groups of local transformations. The internal transformations take place in an abstract space. It has been assumed that the generators of the internal and spacetime groups commute. We proved that the locally ``internal'' groups $SU(2)$ and $U(1)$ are isomorphic to locally ``spacetime'' groups of transformations, and therefore the aforementioned assumption is not true simply because local Lorentz transformations do not commute in general. It has also been assumed that there are no finite dimensional unitary representations of non-compact groups. Since LB1 groups (local boosts plus discrete transformations) are isomorphic to LB2 groups (local spatial rotations) via the compact $SU(2) \times U(1)$, then this is not true either.

\item In the first paper \cite{A} we proved that the group $U(1)$ is isomorphic to the local group of boosts plus discrete transformations on blade one that we called LB1. As the same group $U(1)$ is isomorphic to $SO(2)$, that we also called LB2 since it is related to local tetrad rotations on blade two, then the group $SO(2)$ is isomorphic to the proper group on blade one plus discrete transformations. This is a fundamental result in group theory alone, let alone in physics. We are simultaneously proving, and this is the point that we are emphasizing in this item, that there is an isomorphism between kinematic states and gauge states of the gravitational fields locally. In our present paper we proved two new theorems. First, the local group of $SU(2)$ gauge transformations is isomorphic to the tensor product of three LB1 groups. Second, the local group of $SU(2)$ gauge transformations is isomorphic to the tensor product of three LB2 or $SO(2)$ groups. Then, the local compact $SU(2)$ is isomorphic to the tensor product of three local non-compact LB1 groups (boosts plus discrete transformations). This is another fundamental result in group theory, in physics as well. As in the Abelian case discussed in \cite{A}, and again this is the point that we are emphasizing in this item, we proved again in this non-Abelian case that there is an isomorphism between kinematic states and gauge states of the gravitational fields locally.

\end{enumerate}

The question stands about the possibility of extending these constructions to other field structures that involve other irreducible representations of the Lorentz group. The procedure to build the tetrads, has been laid out in a way that automatically allows for its extension, for instance, to $SU(3)$ gauge theories in a curved spacetime that will be explicitly developed in an upcoming paper. There is an underlying program in these ideas. We quote from \cite{WE}, ``at one time it was even hoped that the rest of physics could be brought into a geometric formulation, but this hope has met with dissapointment, and the geometric interpretation of the theory of gravitation has dwindled to a mere analogy, which lingers in our language in terms like ``metric'', ``affine connection'', and ``curvature'', but is not otherwise very useful''. We rewrote the usual theory of non-Abelian fields in curved spacetime in a new language. The new tetrads replace the standard variables. The gauge transformations of the potentials are reinterpreted through local group isomorphisms as tetrad Lorentz transformations on both blades or planes. In fact this is the way to reexpress the standard gauge theories into a geometrical Riemannian language. The goal is to show that there is a whole new geometrical way to understand or interpret the gauge theories, the equations involved and their solutions under a whole new light. By establishing a link between the local gauge groups of transformations and local geometrical groups of transformations, like in \cite{A}, or in the present manuscript, we are trying to bring the gauge theories into a geometric formulation. The geometrization of the gauge theories is where we are aiming at.

\section{Appendix I}
\label{sec:appI}

The first appendix is telling us that the geodesics through the origin of the $SU(2)$ $2\pi$ parameter sphere, generate a set of tetrad transformations that does not belong to a subgroup of LB1. This is of fundamental importance to prove our theorems. Following the notation in \cite{RG} we write the elements in $SU(2)$ as,

\begin{eqnarray}
S = \sigma_{o}\:\cos(\theta/2) + \imath \: \sigma_{j}\:\hat{\theta}^{j}\:\sin(\theta/2) = \sum\:{1 \over n!}
\left( \sum_{i=1}^{3} {\imath \over 2 } \sigma_{i}\: \theta^{i}\right)^{n}\ ,\label{S}
\end{eqnarray}

where $\sigma_{o}$ is the identity, $\sigma_{j}$ for $j=1\ldots3$ are the usual Pauli matrices, and the summation convention is applied for $j=1\ldots3$. We can then define the function $\theta$ as,

\begin{eqnarray}
\left( \sum_{i=1}^{3} {\imath \over 2 } \sigma_{i}\: \theta^{i}\right)^2 = - \sigma_{o}\: \left( {\theta \over 2} \right)^2 \ , \label{theta}
\end{eqnarray}

\begin{equation}
\theta^2 = \sum_{i=1}^{3}\:\left( \theta^{i}\right)^2\ ,\:\:\:\:\: \mbox{where $\hat{\theta}^{i}$ is given by,} \:\:\:\:\:\hat{\theta}^{i} = \theta^{i} / \mid \theta \mid \ .\label{thetahat}
\end{equation}

Let us consider the $SU(2)$ $2\pi$ parameter sphere in $\theta$ \cite{RG}, and let us evaluate $S$, and $\partial_{\lambda}S$ at $\theta = 0$.
$S$ is just $\sigma_{o}$ at $\theta = 0$. We can write the derivative $\partial_{\lambda}S$ as,

\begin{equation}
\partial_{\lambda}S_{\mid_{\theta=0}} = \left[(-1/2)\:\sigma_{o}\:\sin(\theta/2)\:\partial_{\lambda}\theta + (\imath/2)\:\sigma_{j}\hat{\theta}^{j}\:\cos(\theta/2)\:\partial_{\lambda}\theta + \imath\:\sigma_{j}\:\partial_{\lambda}\hat{\theta}^{j}\:\sin(\theta/2)\right] _{\mid_{\theta=0}} \ .\label{partialS}
\end{equation}

If we consider for instance, all possible geodesics through the $2\pi$ parameter sphere origin, then we must conclude that the four components of $\partial_{\lambda}\theta_{\mid_{\theta=0}}$ can take on any value, ranging from $-\infty$ to $+\infty$. Then, accordingly, the vector components of $Tr[\tilde{\Lambda}^{\alpha}_{\:\:\:\delta}\:\tilde{\Lambda}^{\beta}_{\:\:\:\gamma}\:\Sigma^{\delta\gamma}\:E_{\alpha}^{\:\:\rho}\: E_{\beta}^{\:\:\lambda}\:\ast \xi_{\rho\sigma}\:\ast \xi_{\lambda\tau}\:\partial^{\tau}\:(S)\:S^{-1}]$, can take on any values ranging again, from $-\infty$ to $+\infty$. Borrowing once more the notation and line of thinking from \cite{A}, specially the section gauge geometry, we can see that in correspondence to the scalars we named $C$ and $D$ in \cite{A} we get the scalars $C^{'}$ and $D^{'}$.

Since the vector components of $Tr[\tilde{\Lambda}^{\alpha}_{\:\:\:\delta}\:\tilde{\Lambda}^{\beta}_{\:\:\:\gamma}\:\Sigma^{\delta\gamma}\:E_{\alpha}^{\:\:\rho}\: E_{\beta}^{\:\:\lambda}\:\ast \xi_{\rho\sigma}\:\ast \xi_{\lambda\tau}\:\partial^{\tau}\:(S)\:S^{-1}]$, can take on any values, positive or negative, then we must conclude that $1+C^{'}$ and $D^{'}$ can take on any possible real values. Borrowing again the ideas from \cite{A}, we can analyze as an example, the case where $1+C^{'} > D^{'} > 0$, and $0 > C^{'} > -1$. Let us suppose in addition that $\partial_{\rho}\theta$, $\partial_{\rho}\hat{\theta}^{i}$ and $\hat{\theta}^{i}$ have finite components at the origin. We can always consider the geodesic through the origin of the $2\pi$ sphere, such that $\theta^{i}_{n} = n\: \theta^{i}$, where $n$ is a natural number. Now, $\theta_{n} = n\:\theta$ and $\hat{\theta}^{i}_{n} = \hat{\theta}^{i}$, but $\partial_{\rho}\theta_{n} = n\:\partial_{\rho}\theta$. Then, at the origin of the parameter sphere, $\theta^{i} = 0$, $\theta_{n} = 0$ and $\partial_{\rho}\theta_{n\mid\theta = 0} = n\:\partial_{\rho}\theta\mid_{\theta=0}$. Putting all this toghether we have that for the new geodesic, for $n$ sufficiently large, $D^{'}_{n} > 0 > 1+C^{'}_{n}$. Similar line of thinking for the other cases. This implies one important conclusion. The $SU(2)$ group of local gauge transformations, generates proper and improper LB1 transformations. Therefore the image of $SU(2)$ is not associated to a subgroup of LB1 (tensor products of LB1).

\section{Appendix II}
\label{sec:appII}

The second appendix is introducing the object $\Sigma^{\alpha\beta}$. This object according to the matrix definitions introduced in the references is Hermitic. The use of this object in the construction of our tetrads allows for the local $SU(2)$ gauge transformations $S$, to get in turn transformed into purely geometrical transformations. That is, local rotations of the $U(1)$ electromagnetic tetrads.
The object $\sigma^{\alpha\beta}$ is defined as $\sigma^{\alpha\beta} = \sigma_{+}^{\alpha}\:\sigma_{-}^{\beta}-\sigma_{+}^{\beta}\:\sigma_{-}^{\alpha}$, \cite{MK}$^{,}$\cite{GM}. The object $\sigma_{\pm}^{\alpha}$ arises when building the Weyl representation for left handed and right handed spinors. According to \cite{GM}, it is defined as $\sigma_{\pm}^{\alpha} = (\bf{1},\pm\sigma^{i})$, where $\sigma^{i}$ are the Pauli matrices for $i = 1\cdots3$. Under the $(\frac{1}{2},0)$ and $(0,\frac{1}{2})$ spinor representations of the Lorentz group it transforms as,

\begin{equation}
S_{(1/2)}^{-1}\:\sigma_{\pm}^{\alpha}\:S_{(1/2)} = \Lambda^{\alpha}_{\:\:\:\gamma}\:\sigma_{\pm}^{\gamma}\ .\label{sigmatr}
\end{equation}

Equation (\ref{sigmatr}) means that under the spinor representation of the Lorentz group, $\sigma_{\pm}^{\alpha}$ transform as vectors. In (\ref{sigmatr}), the matrices $S_{(1/2)}$ are local, as well as $\Lambda^{\alpha}_{\:\:\:\gamma}$ \cite{GM}. The $SU(2)$ elements can be considered to belong to the Weyl spinor representation of the Lorentz group. Since the group $SU(2)$ has a homomorphic relationship to $SO(3)$, they just represent local space rotations. It is also possible to define the object $\sigma^{\dagger\alpha\beta} = \sigma_{-}^{\alpha}\:\sigma_{+}^{\beta}-\sigma_{-}^{\beta}\:\sigma_{+}^{\alpha}$, analogously. Then, we have,

\begin{center}
$\imath \: \left(\sigma^{\alpha\beta} + \sigma^{\dagger\alpha\beta}  \right)  = \left\{ \begin{array}{ll}
				0 \:\:\:\:\: \mbox{if $\alpha = 0$ and $\beta = i$}\\
				4\:\epsilon^{ijk}\:\sigma^{k} \:\:\:\:\: \mbox{if $\alpha = i$ and $\beta = j$ \ ,}
				    \end{array}
			    \right. $
\end{center}

\begin{center}
$ \sigma^{\alpha\beta} - \sigma^{\dagger\alpha\beta}  = \left\{ \begin{array}{ll}
				-4\:\sigma^{i} \:\:\:\:\: \mbox{if $\alpha = 0$ and $\beta = i$}\\
				0 \:\:\:\:\: \mbox{if $\alpha = i$ and $\beta = j$ \ .}
				    \end{array}
			    \right. $
\end{center}

We might then call $\Sigma_{ROT}^{\alpha\beta} = \imath \: \left(\sigma^{\alpha\beta} + \sigma^{\dagger\alpha\beta} \right)$, and $\Sigma_{BOOST}^{\alpha\beta} = \imath \: \left(\sigma^{\alpha\beta} - \sigma^{\dagger\alpha\beta} \right)$. Therefore, a possible choice for the object $\Sigma^{\alpha\beta}$ could be for instance $\Sigma^{\alpha\beta} = \Sigma_{ROT}^{\alpha\beta} + \Sigma_{BOOST}^{\alpha\beta}$. This is a particularly suitable choice when we consider proper Lorentz transformations of the tetrad vectors nested within the structure of the gauge vectors $X^{\mu}$ and $Y^{\mu}$. For spatial, that is, rotations of the $U(1)$ electromagnetic tetrad vectors which in turn are nested within the structure of the two gauge vectors $X^{\mu}$ and $Y^{\mu}$, as is the case under study in this paper, we can simply consider $\Sigma^{\alpha\beta} = \Sigma_{ROT}^{\alpha\beta}$. These possible choices also ensure the Hermiticity of gauge vectors.  Since in the definition of the gauge vectors $X^{\mu}$ and $Y^{\mu}$ we are taking the trace, then $X^{\mu}$ and $Y^{\mu}$ are real. All the greek indices $\alpha$, $\beta$, $\delta$, $\epsilon$, $\gamma$, and $\kappa$, have been reserved in this manuscript for locally inertial coordinate systems. These $\Sigma^{\alpha\beta}$ objects are specially introduced, and its importance described in the section ``gauge geometry'' (\ref{gaugegeometry}) specifically when we say:  ``We observe also the following. Since locally inertial coordinate transformations in general do not commute, then the locally $SU(2)$ generated transformations are non-Abelian. The non-Abelianity of $SU(2)$ is mirrored by the non-commutativity of these locally inertial coordinate transformations $\tilde{\Lambda}^{\alpha}_{\:\:\:\delta}$ of the electromagnetic tetrads. The key role in this non-commutativity is played by the object $\Sigma^{\alpha\beta}$, that translates local $SU(2)$ gauge transformations, into locally inertial Lorentz transformations''.

\section{Appendix III}
\label{sec:appIII}

Let us suppose the following hypothesis. We consider two $SU(2)$ gauge transformations $S_{1}$ and $S_{2}^{-1}$ that when composed generate no LB1 transformation but do generate when composed, a non-trivial spatial transformation of the electromagnetic tetrad. That is, $\tilde{\Lambda}^{\alpha}_{1\:\:\delta}$ and $\tilde{\Lambda}^{\alpha}_{2\:\:\delta}$ are not the same, but since the composition generates no LB1 transformation, then $(S_{2}^{-1}\:S_{1})^{-1}\:\partial_{\mu}(S_{2}^{-1}\:S_{1}) = 0$. This last equation implies that $S_{1}= S_{g}\:S_{2}$, where $S_{g}$ is any constant or global $SU(2)$ gauge transformation. Therefore all our conclusions about group isomorphisms apply to equivalence classes of $SU(2)$ gauge transformations. We say that if $S$ is a local $SU(2)$ transformation, then the class of equivalence is made up of all the products $S_{g}\:S$. Therefore, the theorems that we proved are valid for these classes of equivalence. This is analogous to adding constants to scalars $\Lambda$ in the Abelian case \cite{A}. Another point that we did not discuss is why locally $SU(2)$ generated LB1 or LB2 transformations commute with locally $U(1)$ generated LB1 or LB2 transformations. The answer is simply because if we add two gauge vectors $X_{1}^{\mu} + X_{2}^{\mu}$, the LB1 or LB2 generated gauge transformations are going to commute either on blade one or two. In particular if the $SU(2)$ gauge vector is $X_{1}^{\mu} = Tr[\Sigma^{\alpha\beta}\:E_{\alpha}^{\:\:\rho}\: E_{\beta}^{\:\:\lambda}\:\ast \xi_{\rho}^{\:\:\sigma}\:\ast \xi_{\lambda\tau}\:A^{\tau}]$ and the $U(1)$ gauge vector is $X_{2}^{\mu} = A^{\mu}$, the gauge transformations generated by both gauge vectors are going to commute, either on blade one, or blade two. We remind ourselves from section \ref{gaugegeometry} that the $SU(2)$ ``gauge vector'' $X_{1}^{\mu} = Tr[\Sigma^{\alpha\beta}\:E_{\alpha}^{\:\:\rho}\: E_{\beta}^{\:\:\lambda}\:\ast \xi_{\rho}^{\:\:\sigma}\:\ast \xi_{\lambda\tau}\:A^{\tau}]$ was built to be $U(1)$ gauge invariant. Finally, we would like to make one more issue clear, regarding the theorems that we proved. When we talk about the local gauge group $SU(2)$ we are talking either of the connected component or of the disconnected component. Since the connected component and the disconnected component are isomorphic to each other, then, the group itself is isomorphic to the tensor product of three LB1 and three LB2 groups, for instance.

\acknowledgements

%I am grateful to R. Gambini and J. Pullin for reading the original version of this manuscript and
%for many fruitful discussions.
This work was partially funded by PEDECIBA.

%\bibliography{your-bib-file} % place the references here.

\end{document}